\documentstyle[twocolumn,prl,aps,epsfig,amssymb]{revtex}
\tighten
\let\jnfont=\rm
\def\NPB#1,{{\jnfont Nucl.\ Phys.\ B }{\bf #1},}
\def\PLB#1,{{\jnfont Phys.\ Lett.\ B }{\bf #1},}
\def\EPJC#1,{{\jnfont Eur.\ Phys.\ Jour.\ C }{\bf #1},}
\def\PRD#1,{{\jnfont Phys.\ Rev.\ D }{\bf #1},}
\def\PRL#1,{{\jnfont Phys.\ Rev.\ Lett.\ }{\bf #1},}
\def\MPLA#1,{{\jnfont Mod.\ Phys.\ Lett.\ A }{\bf #1},}
\def\JPG#1,{{\jnfont J.\ Phys.\ G}{\bf #1},}
\def\CTP#1,{{\jnfont Commun.\ Theor.\ Phys.\ }{\bf #1},}

\begin{document}
\draft
\preprint{}

\title{ Lightest Higgs Boson Mass in Split Supersymmetry with See-saw Mechanism}

\author{Junjie Cao$^{1,2}$,  Jin Min Yang$^{3,2}$}

\address{ \ \\[2mm]
{\it $^1$ Department of Physics, Henan Normal University, Xinxiang, Henan 453002, China;}\\ [2mm]
{\it $^2$ Institute of Theoretical Physics, Academia Sinica, Beijing 100080, China;} \\ [2mm]
{\it $^3$ CCAST(World Laboratory), P.O.Box 8730, Beijing 100080, China;} \ \\[6mm] }

\date{\today}

\maketitle

\begin{abstract}
In the minimal supersymmetric standard model extended by 
including right-handed neutrinos with see-saw mechanism,
the neutrino Yukaka couplings can be as large as the top-quark 
Yukawa couplings and thus the neutrino/sneutrino may cause sizable 
effects in Higgs boson self-energy loops. Our explicit one-loop calculations
show that the neutrino/sneutrino effects may have an opposite sign
to top/stop effects and thus lighten the lightest Higgs boson. If
the soft-breaking mass of the right-handed neutrino is very large
(at the order of Majorana mass scale), such as in the split-SUSY
scenario, the effects can lower the lightest Higgs boson mass by a
few tens of GeV.  So the Higgs mass bound of about 150 GeV 
in split-SUSY may be lowered significantly if right-handed neutrinos 
come into play with see-saw mechanism.

\end{abstract}


Searching for the Higgs boson is a primary goal of ongoing and
forthcoming high energy colliders. The verification or disproof of
the existence of a light Higgs boson will serve as an acid test
for both the Standard Model (SM) and the supersymmetry (SUSY)
extensions. While for the SM a light Higgs boson is favored
indirectly by the precision electroweak data, the Minimal
Supersymmetric Standard Model (MSSM) predicts a Higgs boson ($h$)
lighter than about 130 GeV \cite{h-original,h-limit1,h-limit2}. In
the recently proposed split-SUSY scenario \cite{split},  where the
fine-tuning problem is argued not to be solved by SUSY and thus
the soft-breaking masses of sfermions can be very large,  such a
bound is relaxed to about 150 GeV. It is well known that these
bounds, much higher than the tree-level prediction $m_h<m_Z$, are
obtained by considering loop effects, mainly the top/stop loop
effects. The underlying reason for the sizable top/stop loop
effects is the largeness of the top-quark Yukawa couplings. So, if
any other particles have large Yukawa couplings, their loop
effects in the Higgs boson mass should also be taken into account.

The recent compelling evidence of neutrino oscillation indicates
massive neutrinos, which necessitates the introduction of right-handed
neutrino superfields ~\cite{Hisano} in the MSSM. If the Majorana
masses of the right-handed neutrinos are large enough, the tiny
masses of light neutrinos can be naturally obtained, which is the
so-called see-saw mechanism~\cite{Yanagida}. In such a
supersymmetric see-saw model the neutrino Yukaka couplings can be
as large as the top-quark Yukawa couplings \cite{Wyler}. Thus, the
neutrino/sneutrino may cause sizable loop effects in Higgs boson
masses. In this note we focus on such effects and perform an
explicit one-loop calculations \footnote{Note that there are two approaches
for calculating such leading loop effects. One is the explicit loop calculations 
used in this work. The other is the renormalization group technique. We note that
in the approach of explicit loop calculations, one-loop calculations may be not 
accurate enough and thus higher order loop effects may need to be considered.
But in this letter we only perform one-loop calculations in order to illustrate 
the possible size of the effects. In our future work we will provide more
complete calculations, either by considering higher order loops or by the 
renormalization group to resum the leading logs.}.

We start our calculation from the superpotential.
Compared with the MSSM, the supersymmetric see-saw model contains the
additional right-handed neutrino superfield in the superpotential
\begin{eqnarray}
- \frac{1}{2} \nu_R^{c} M_R \nu_R^c + \nu_R^{c} {\bf Y_\nu} L\cdot H_2 \ ,
\end{eqnarray}
where $Y_{\nu}$ is the Yukawa couplings of neutrinos and $M_R$ is
the Majorana mass, both of which are generally $3\times 3$
matrices in the generation space. For simplicity and illustration, we
in our analysis ignore the flavor structure and consider only one
generation of neutrinos \footnote{The general result with the three
families of neutrinos is quite complicated and will be presented
elsewhere \cite{cao-phd}.}. Such an addition of right-handed
neutrino superfields will enrich the phenomenology of both
neutrino sector and sneutrino sector.

First, we examine the neutrino sector. The neutrino mass term is
given by
\begin{eqnarray}
& &(\nu_L, \nu_R^c)
    \left ( \begin{array}{cc} 0 & Y_\nu v_2 \\
             Y_{\nu} v_2 & M_R\end{array} \right )
  \left ( \begin{array}{c} \nu_L \\ \nu_R^c \end{array} \right ) + h.c. \nonumber\\
& &=-(\nu_1, \nu_2)
    \left ( \begin{array}{cc} \frac{Y_\nu^2 v_2^2}{M_R} & 0 \\
             0 & M_R ( 1 + \frac{Y_\nu^2 v_2^2}{M_R^2} ) \end{array} \right )
  \left ( \begin{array}{c} \nu_1 \\ \nu_2 \end{array} \right ) + h.c. \ .
\end{eqnarray}
Here $v_2$ is the vacuum expectation value of the Higgs doublet $H_2$ and
the mass matrix is diagonalized by an unitary rotation, which
rotates $\nu_L$ and $\nu_R^c$ into two mass eigenstates
$\nu_{1,2}$, i.e.,
\begin{eqnarray}
 \left ( \begin{array}{c} \nu_1 \\ \nu_2 \end{array} \right )
  = \left ( \begin{array}{cc} \cos\theta_\nu & \sin\theta_\nu \\
                             -i \sin\theta_\nu & i \cos\theta_\nu  \end{array} \right )
  \left ( \begin{array}{c} \nu_L \\ \nu_R^c \end{array} \right )
\end{eqnarray}
with the mixing angle $\theta_\nu$ determined by
\begin{eqnarray}
\tan(2\theta_\nu)=-\frac{2Y_\nu v_2}{M_R}\ .
\end{eqnarray}
 Note that all these states are the two-component spinors.
We can define the four-component Majorana spinors as $\nu_1 \equiv (\nu_1, \bar\nu_1)^T$
and $\nu_2 \equiv (\nu_2, \bar\nu_2)^T$, which will be used in our following calculations.

Then the neutrino Yukawa couplings with neutral Higgs bosons are given by
\begin{eqnarray}
&& \frac{1}{2\sqrt{2}} \left[
    (H \sin \alpha + h \cos \alpha ) \bar{\nu}_i ( \xi_{ij} P_L + \xi^\dag_{ij} P_R ) \nu_j
   \right. \nonumber \\
&&~~~~~~~~~~~~~~ \left. +i \cos\beta  A \bar{\nu}_i ( \xi_{ij}
P_L-\xi^\dag_{ij} P_R ) \nu_j \right ]\ ,
\end{eqnarray}
where $\alpha$ is the mixing angle between the two neutral CP-even Higgs bosons,
$\tan\beta=v_2/v_1$ is the ratio of the vacuum expectation values of the
two Higgs doublets, $P_{L,R}\equiv (1\mp \gamma_5)/2$,
and $\xi_{ij}$ are given by
\begin{eqnarray}
\xi= \left ( \begin{array}{cc} \sin 2 \theta_\nu Y_\nu &  -i \cos 2
\theta_\nu Y_\nu \\ - i \cos 2 \theta_\nu Y_\nu & \sin 2
\theta_\nu Y_\nu \end{array} \right ) \ .
\end{eqnarray}
Now we turn to the sneutrino sector. The sneutrino mass terms arise from
F-terms, D-terms as well as the soft-breaking terms given by
\begin{eqnarray}
V_{soft}&= &  \tilde{L}^\ast m_{\tilde{L}}^2 \tilde{L}
   +\tilde{\nu}_R^{c \ast} m_{\tilde{\nu}_R}^2  \tilde{\nu}_R^{c} \nonumber \\
& &   + B^\ast \tilde{\nu}_R^{c \ast} M_R^\ast
   \tilde{\nu}_R^{c \ast} + B \tilde{\nu}_R^{c} M_R \tilde{\nu}_R^{c}  \nonumber \\
& &   - A_\nu \epsilon_{\alpha \beta} H_2^\alpha \tilde{\nu}_R^c
Y_\nu
   \tilde{L}^\beta
  - A_\nu^\ast \epsilon_{\alpha \beta} H_2^{\ast \alpha}
       \tilde{\nu}_R^{c \ast} Y_\nu^\ast \tilde{L}^{\ast \beta}
      \ .
\end{eqnarray}
Throughout this paper, we assume $M_R > B, A_{\nu}$ and $M_R \gg v_2, m_Z$, and treat
the soft terms proportional to $B$ and $A_{\nu}$ as interactions. Then the sneutrino mass terms are
given by \cite{Haber}
\begin{eqnarray}
&& \left ( \begin{array}{cc}\tilde{\nu}_L^\ast, \tilde{\nu}_R^{c\ast}
                     \end{array} \right )
   \left (\begin{array}{cc}
   \tilde M_{LL}^2  &  M_R Y_\nu v_2 \\
    M_R Y_\nu v_2 & \tilde M_{RR}^2 \end{array} \right )
   \left (\begin{array}{c} \tilde{\nu}_L \\ \tilde{\nu}_R^c  \end{array} \right )
   \nonumber \\
&&\approx
 \left ( \begin{array}{cc}\tilde \nu_1^\ast, \tilde \nu_2^\ast
                     \end{array} \right )
      \left (\begin{array}{cc}
    m_{\tilde{L}}^2 & 0 \\
    0 & M_R^2 + m_{\tilde{\nu}_R}^2  \end{array} \right )
   \left (\begin{array}{c} \tilde \nu_1 \\ \tilde \nu_2 \end{array} \right ) \ ,
\label{snu}
\end{eqnarray}
where
\begin{eqnarray}
\tilde M_{LL}^2 & =& m_{\tilde{L}}^2 + Y_\nu^2 v_2^2 + \frac{1}{2} m_Z^2 \cos 2 \beta \ ,\\
\tilde M_{RR}^2 & =& M_R^2 + Y_\nu^2 v_2^2 + m_{\tilde{\nu}_R}^2 \ .
\end{eqnarray}
The sneutrino mass matrix is diagonalized by an unitary rotation in Eq.(\ref{snu})
and the two mass eigenstates $\tilde \nu_{1,2}$ are given by
\begin{eqnarray}
\left (\begin{array}{c} \tilde \nu_1 \\ \tilde \nu_2 \end{array} \right )
 = \left ( \begin{array}{cc}\cos\theta_{\tilde\nu}&\sin\theta_{\tilde\nu} \\
                            -\sin\theta_{\tilde\nu} &\cos\theta_{\tilde\nu}   \end{array} \right )
\left (\begin{array}{c} \tilde \nu_L \\ \tilde \nu_R^c \end{array} \right )
\end{eqnarray}
with the mixing angle $\theta_{\tilde{\nu}}$ determined by
\begin{eqnarray}
\sin (2 \theta_{\tilde{\nu}}) = \frac{2 Y_\nu M_R v_2}
                              { m_{\tilde\nu_1}^2 - m_{\tilde\nu_2}^2} \ .
\end{eqnarray}
Then the sneutrino Yukawa couplings are given by
\begin{eqnarray}
-\frac{Y_\nu}{\sqrt{2}}&& \left \{ (2 Y_\nu v_2 + \sin 2
\theta_{\tilde{\nu}} M_R )
    (H \sin \alpha + h \cos \alpha )\tilde{\nu}_1^\ast \tilde{\nu}_1 \right .\nonumber \\
&& +(2 Y_\nu v_2 - \sin 2 \theta_{\tilde{\nu}} M_R )
    (H \sin \alpha + h \cos \alpha ) \tilde{\nu}_2^\ast \tilde{\nu}_2   \nonumber \\
&& +\cos 2 \theta_{\tilde{\nu}} M_R
    (H \sin \alpha + h \cos \alpha )
   (\tilde{\nu}_1^\ast \tilde{\nu}_2 +\tilde{\nu}_2^\ast \tilde{\nu}_1) \nonumber \\
&& - i M_R \cos \beta  A
    (\tilde{\nu}_1^\ast \tilde{\nu}_2-\tilde{\nu}_2^\ast \tilde{\nu}_1) \nonumber \\
&& + \frac{Y_\nu}{\sqrt{2}} \left [ (H \sin \alpha + h \cos \alpha
)^2 \right.
                                                              \nonumber \\
&&\left. \left. +A^2 \cos^2 \beta \right ]
      (\tilde{\nu}_1^\ast \tilde{\nu}_1+ \tilde{\nu}_2^\ast \tilde{\nu}_2)  \right \} \ .
\label{interaction}
\end{eqnarray}
In addition, the bilinear and trilinear interactions from the soft-breaking
terms are given by
\begin{eqnarray}
&&  -B (\sin \theta_{\tilde{\nu}} \tilde{\nu}_1 +  \cos
\theta_{\tilde{\nu}} \tilde{\nu}_2 ) M_R (\sin
\theta_{\tilde{\nu}} \tilde{\nu}_1 +  \cos \theta_{\tilde{\nu}}
\tilde{\nu}_2 )  \nonumber \\
&& - A_{\nu} Y_\nu (\sin \theta_{\tilde{\nu}}
\tilde{\nu}_1 +  \cos \theta_{\tilde{\nu}} \tilde{\nu}_2)
(\cos \theta_{\tilde{\nu}} \tilde{\nu}_1 -  \sin\theta_{\tilde{\nu}}
\tilde{\nu}_2 ) \nonumber \\
&&   \times \left ( v_2 + \frac{1}{\sqrt{2}}  ( H \sin \alpha + h
\cos\alpha + i A \cos \beta ) \right ) +h.c.
\end{eqnarray}
The neutrino/sneutrino contribute to Higgs boson self-energies
through the diagrams shown in Fig.1. The renormalized
self-energies are given by
\begin{eqnarray}
\hat{\Sigma}_h&\approx& -\omega_\nu \cos^2 \alpha \ ,\\
\hat{\Sigma}_H&\approx & - \omega_\nu \sin^2 \alpha \ ,\\
\hat{\Sigma}_h H&\approx & - \omega_\nu \sin \alpha \cos \alpha \ ,
\end{eqnarray}
where $\omega_\nu$  can be decomposed as
\begin{eqnarray}
\omega_\nu = \omega_{\nu}^{SUSY} +\omega_{\nu}^{soft}
\end{eqnarray}
with $\omega_{\nu}^{SUSY}$ and $\omega_{\nu}^{soft}$ representing
respectively the contribution from supersymmetric part and soft-breaking part
given by
\begin{eqnarray}
 \omega_\nu^{SUSY}&=& \frac{Y_\nu^4 v_2^2}{4 \pi^2} \left [
   \frac{1}{2} \ln\frac{m_{\tilde{\nu}_1}^2 m_{\tilde{\nu}_2}^2}{M_R^4}
  -1  \right . \nonumber \\
&& + \frac{ M_R^2}{m_{\tilde{\nu}_1}^2-m_{\tilde{\nu}_2}^2}
                    \ln\frac{m_{\tilde{\nu}_1}^2}{m_{\tilde{\nu}_2}^2}
   + \frac{M_R^4}{(m_{\tilde{\nu}_1}^2-m_{\tilde{\nu}_2}^2)^2}  \nonumber \\
&&\left. \times \left ( 1 -\frac{m_{\tilde{\nu}_1}^2
  +m_{\tilde{\nu}_2}^2}{2 (m_{\tilde{\nu}_1}^2 - m_{\tilde{\nu}_2}^2)}
\ln\frac{m_{\tilde{\nu}_1}^2}{m_{\tilde{\nu}_2}^2} \right ) \right ] \ ,\\
\omega^{soft}_\nu&=& -\frac{Y_\nu^2}{16 \pi^2}\frac{B^2
M_R^2}{m_{\tilde{\nu}_2}^2}
  \left [  \frac{4}{3 m_{\tilde{\nu}_2}^2}
           (2 Y_\nu v_2 - \sin 2 \theta_{\tilde{\nu}} M_R)^2
          \right. \nonumber \\
&& -2\sin^2 2 \theta_{\tilde\nu} \frac{M_R^2}
         {(m_{\tilde{\nu}_2}^2-m_{\tilde{\nu}_1}^2)^3}  \nonumber \\
&&\left. \times \left (m_{\tilde{\nu}_2}^4
         -m_{\tilde{\nu}_1}^4 -2 m_{\tilde{\nu}_2}^2 m_{\tilde{\nu}_1}^2
          \ln \frac{m_{\tilde{\nu}_1}^2}{m_{\tilde{\nu}_2}^2} \right ) \right ]
          \nonumber \\
&&- \frac{Y_\nu^2}{16 \pi^2} \frac{ M_R^2}
                 {(m_{\tilde{\nu}_2}^2 -m_{\tilde{\nu}_1}^2)^3}
 (2 B^2 M_R^2 \sin^2 2 \theta_{\tilde{\nu}}
         \nonumber \\
&& + 4 B M_R A_\nu v_2 \sin 2 \theta_{\tilde{\nu}}
   +A_\nu^2 v_2^2) \nonumber \\
&& \times \left[ 2 (m_{\tilde{\nu}_1}^2-m_{\tilde{\nu}_2}^2)
       -(m_{\tilde{\nu}_1}^2+m_{\tilde{\nu}_2}^2)
       \ln \frac{m_{\tilde{\nu}_1}^2}{m_{\tilde{\nu}_2}^2} \right]
    \label{omega}
\end{eqnarray}
\begin{figure}[]
\hspace*{-1cm} \epsfig{file=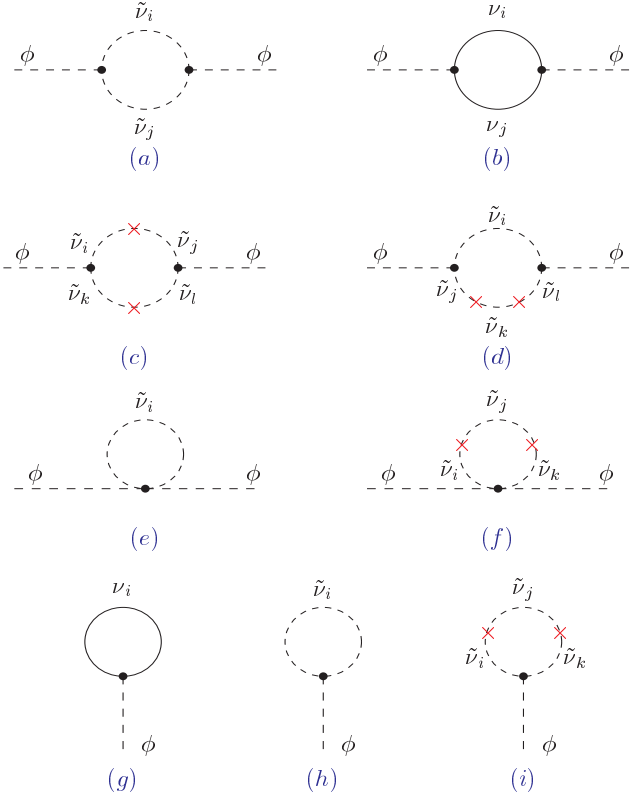,width=9cm, height=11cm}
\caption{Feynman diagrams of neutrino/sneutrino effects in Higgs
boson self-energies. $\phi$ denotes a neutral Higgs boson, i.e., $h$, $H$ or $A$.
The symbol '$\times$' means a soft-term mass insertion.}
\end{figure}
In the calculation of $\omega_{\nu}^{soft}$, we dropped the terms
suppressed by $v_2^2/m_{\tilde{\nu}_R}^2$. We also checked that
the contributions from higher order mass insertions are
negligible.

The lightest Higgs boson mass with one-loop
neutrino/sneutrino effects is given by\cite{Dabelstein}
\begin{eqnarray}
 &&  m^2_h =  \frac{M_A^2 + M_Z^2 + \omega_\nu} {2}
     -\left[ \frac{ (M_A^2 + M_Z^2)^2 + \omega^2_\nu}{4} \right. \nonumber \\
 &&~~~~ \left. -M_A^2 M_Z^2 \cos^2 2\beta
        + \frac{\omega_\nu \cos 2\beta}{2}  (M_A^2 - M_Z^2)
        \right]^{1/2}
\end{eqnarray}
From Eq.(\ref{omega}) and Eq.(\ref{snu}), we see that the relevant
SUSY parameters are $A_\nu$,$B$,$m_{\tilde L}$,
$m_{\tilde{\nu}_R}$ and $M_R$. $M_R$ must be much larger than
$m_Z$ in order to drive the see-saw mechanism, but the scales of
the soft-breaking parameters are not clear\footnote{Note that
$M_R$, $m_{\tilde\nu_R}$ and $B$ are associated with the $SU(2)\times U(1)$ 
singlet superfield $\nu_R$ and thus may have fundamentally different origin 
from $A_\nu$ and $m_{\tilde L}$.}.
If one insists on naturalness requirement, these soft breaking parameters
should not be much larger than the electroweak scale \cite{B-value}.
In this case, since the SUSY breaking in neutrino/sneutrino sector
is not significant, the neutrino/sneutrino loop effects on 
the Higgs boson mass are small, as will be shown in our numerical results.
In our numerical calculations we abandon the naturalness, as in the
split-SUSY scenario, and thus the soft-breaking masses for sneutrinos can be 
much larger. 

\begin{figure}[]
\hspace*{-0.5cm} \epsfig{file=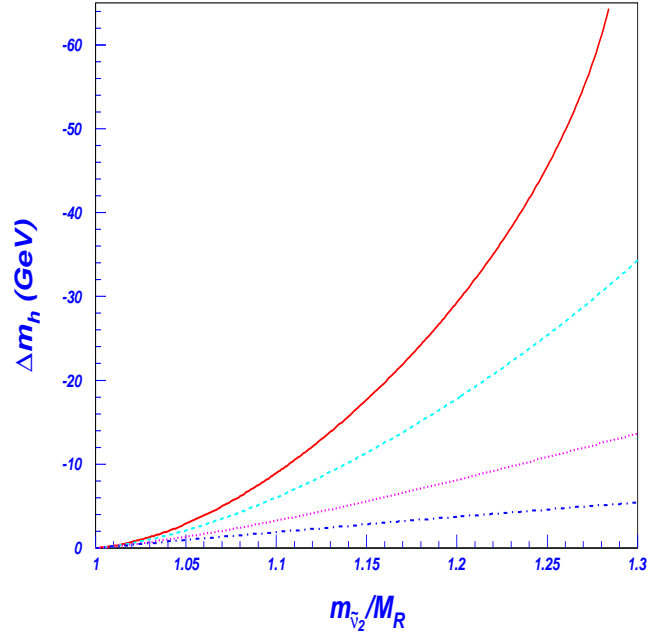,width=8.7cm, height=8.7cm}
\caption{The neutrino/sneutrino contribution to $m_h$: the solid, dashed, dotted and
dash-dotted curve (from up to down) is for $m_{\tilde L}=10^3$, $10^7$, $10^{11}$, $10^{13}$ GeV,
respectively. $M_R=10^{14}$ GeV and other parameters are fixed in the context. }
\end{figure}

In Fig.~2 we show the neutrino/sneutrino contribution to $m_h$,
i.e., $\Delta m_h \equiv m_h -m_h^{\rm tree}$, versus $m_{\tilde
\nu_2}/M_R$ for $M_R=10^{14}$ GeV and different values of $m_{\tilde L}$.
Other parameters are fixed as 
$m_A=200$ GeV, $\tan\beta=30$, $Y_\nu=1$ and $B=A_\nu=0$.
With above choice of parameters, only $\omega_{\nu}^{SUSY}$ contribute to
our numerical results.

Since $m_{\tilde \nu_2}^2\approx M_R^2 + m_{\tilde \nu_R}^2$ and
$m_{\nu_2}\approx M_R$, we see from Fig.~2 that the effects of
neutrino/sneutrino are sensitive to the mass splitting of
$\tilde \nu_2$ (heavy sneutrino) and $\nu_2$ (heavy neutrino).
Obviously, such a mass splitting is determined by the right-handed
neutrino soft-breaking mass $m_{\tilde \nu_R}$. So  the effects of
neutrino/sneutrino depend on the size of the soft masses $m_{\tilde \nu_R}$
and $m_{\tilde L}$.  

Fig.~2 shows that if $m_{\tilde \nu_R}$ is as large as $M_R$, the
effects can be quite significant, lowering $m_h$ by a few tens of
GeV. If $m_{\tilde \nu_R}$ is far smaller than $M_R$, e.g., of the
order of the electroweak scale, then the effects are negligibly
small. This can be understood from $h\tilde{\nu}_1
\tilde{\nu}_1^\ast $ interactions shown in Eq.(\ref{interaction}):
for $m_{\tilde \nu_R} \ll M_R $ we have $m_{\tilde{\nu}_2} \simeq
M_R$ and then the coupling tends to be zero. As a result, the first
term in $\omega_{\nu}^{SUSY}$ is cancelled by other terms. When $
m_{\tilde \nu_R} $ becomes large, such cancellation is alleviated.
Another way to understand the above results is that for soft-breaking
parameters of electroweak scale, the SUSY-breaking in neutrino/sneutrino 
sector is not significant and thus the contributions from neutrino loops 
tend to be cancelled out by those from sneutrino loops.

In case of large SUSY breaking in neutrino/sneutrino sector, the neutrino/sneutrino 
loop effects are proportional to 
$\log(m_{\tilde \nu_1}/m_{\tilde \nu_2}) 
\approx \log(m_{\tilde L}/\sqrt{M_R^2+m_{\tilde \nu_R}^2})$. A smaller $m_{\tilde L}$
leads to a larger splitting between $m_{\tilde \nu_1}$ and $m_{\tilde \nu_2}$ and thus
gives a larger contribution.

We found that the results are not sensitive to $\tan\beta$ and $m_A$.
Note that we fixed $B=A_\nu=0$ and thus $\omega_{\nu}^{soft}$ vanished in Fig.~2.
If we choose a non-zero $B$ or $A_\nu$, we find that the contribution of
$\omega_{\nu}^{soft}$ to $m_h$ is also negative, whose magnitude depends on
the size of $B$ and $A_\nu$. For $B\ll M_R$ and $A_\nu\ll M_R$, the contribution
of $\omega_{\nu}^{soft}$ is negligible. But if any one,  $B$ or $A_\nu$, is
as large as $M_R$,  $\omega_{\nu}^{soft}$ can lower $m_h$ by a few tens of GeV,
regardless of the magnitude of $m_{\tilde \nu_R}$.

Since the effects depend sensitively on the right-handed neutrino soft-breaking
masses, it is important to know how large they possibly are.
Unfortunately, their origin is not clear and thus their values are theroretically
arbitrary so far.  In the split-SUSY scenario \cite{split}, the soft-breaking
masses of sfermions (and thus $m_{\tilde \nu_R}$) can be very large. Therefore,
the neutrino/sneutrino loop effects on Higgs boson mass may be sizable in split-SUSY
if right-handed neutrinos are introduced with see-saw mechanism.

Including such large negative effects, the upper bound on the lightest 
Higgs boson mass of about 150 GeV in split-SUSY \cite{split} will be
lowered significantly. The LEP experiment already set a lower bound of about
90 GeV \cite{LEP-bound} on the SUSY Higg mass and the Fermilab Tevatron collider 
will further push up the lower bound in case of unobservation. So the lightest 
Higgs boson mass will be a crucial test for various SUSY models. 

We note that the large Yukawa couplings of neutrinos or large soft-masses of right-handed
      neutrinos will not cause any problems in the precision electroweak fit of the SM,
      such as the parameterized variables $S$, $T$ and $U$. The reason is that the right-handed
      nuetrinos or sneutrinos have no gauge couplings. The left-handed sneutrinos contribute
      to $S$, $T$ and $U$ in a similar way as other sfermions and their effects decouple as
      they get heavy.

We conclude that due to the unsuppressed neutrino Yukawa couplings
in SUSY see-saw model, the neutrino/sneutrino may cause sizable
effects in the lightest Higgs boson mass. Such effects have an
opposite sign to the top/stop effects and thus lighten the
lightest Higgs boson. If the soft-breaking mass of the
right-handed neutrino is sufficiently large, which can be realized
in the split-SUSY scenario, the effects can lower the mass bound
by a few tens of GeV.

We thank Ken-ichi Hikasa for comments and Mashahiro Yamaguchi for discussions.
This work was supported in part by National Natural Science Foundation of China
(NNSFC).

\end{document}